\documentclass[floatfix,aps,amsmath,nofootinbib,twocolumn,10pt]{revtex4}

\usepackage{listings}
\usepackage{graphicx}
\usepackage{bm}
\usepackage{rotating}
\usepackage{array}
\usepackage{amsmath}
\usepackage{amssymb} 
\usepackage{mathrsfs} 
\usepackage{cancel}
\usepackage{subfig}
\usepackage{float}
\usepackage{caption}
\usepackage{xcolor}

\def\({\left(}
\def\){\right)}
\def\[{\left[}
\def\]{\right]}

\def\e{\begin{equation}}
\def\q{\end{equation}}
\def\m{\begin{eqnarray}}
\def\n{\end{eqnarray}}

\newcommand{\Mov}[1]{{\color{black}{#1}}}
\newcommand{\Ke}[1]{{\color{black}{#1}}}
\newcommand{\Tan}[1]{{\color{black}{#1}}}

\begin{document}
\title{Gravitational Waves from Collisions of Fuzzy Dark Matter Solitons}
\author{Chen Tan$^{1,2,3,4}$}
\thanks{These authors contributed equally to this paper.}
\author{Jing-Kang Bin$^{1,2,3,4}$}
\thanks{These authors contributed equally to this paper.}
\author{Ke Wang$^{5}$}
\thanks{Corresponding author: {wangke@lnnu.edu.cn}}
\affiliation{$^1$School of Physical Science and Technology, Lanzhou University, Lanzhou 730000, China}
\affiliation{$^2$Institute of Theoretical Physics $\&$ Research Center of Gravitation, Lanzhou University, Lanzhou 730000, China}
\affiliation{$^3$Key Laboratory of Quantum Theory and Applications of MoE, Lanzhou University, Lanzhou 730000, China}
\affiliation{$^4$Lanzhou Center for Theoretical Physics $\&$ Key Laboratory of Theoretical Physics of Gansu Province, Lanzhou University, Lanzhou 730000, China}
\affiliation{$^5$Department of Physics, Liaoning Normal University, Dalian 116029, China}

\date{\today}

\begin{abstract}
According to the Schrödinger-Poisson (SP) equations, fuzzy dark matter (FDM) can form a stable equilibrium configuration, the so-called FDM soliton. The SP system can also determine the evolution of FDM solitons, such as head-on collision. In this paper, we first propose a new adimensional unit of length, time and mass. And then, we simulate the adimensional SP system with $\mathtt{PyUltraLight}$ to study the GWs from collision of FDM solitons when the linearized theory is valid and the GW back reaction on the evolution of FDM solitons is ignored. Finally, we find that the GWs from collisions have a frequency of (few ten-years)$^{-1}$ or (few years)$^{-1}$ when FDM mass is $m=10^{-18}\rm{eV}/c^2$ or $m=10^{-17}\rm{eV}/c^2$. Therefore, future detection of such GWs will constrain the property of FDM particle and solitons.
\end{abstract} 

\maketitle


\section{Introduction}
\label{sec:intro}
Gravitational waves (GWs) are ripples in spacetime. Einstein's general relativity predicts that this kind of gravitational radiation involves a spherically or rotationally asymmetric acceleration among the masses. More precisely, an isolated system will radiate GWs when the second (or third, ...) time derivative of the quadrupole (or octupole, ...) moment of its stress–energy tensor is nonzero. The Universe is expected to be populated by GWs spanning orders of magnitude in frequency. For example, $\gtrsim10^{-16}$Hz primordial GWs due to primordial tensor fluctuations during cosmic inflation left their footprints in the B-modes polarization anisotropies of the cosmic microwave background radiation (CMB), which is allegedly detected by CMB polarization telescope~\cite{BICEP:2021xfz,Li:2017drr}; binary supermassive black holes in galactic nuclei emit about $10^{-8}-10^{-3}$Hz GWs, whose low-frequency end is supposed to be detected by pulsar timing arrays (PTA)~\cite{NANOGrav:2023gor,EPTA:2023fyk,Reardon:2023gzh}; compact binaries or black hole binaries in galaxies emit about $10^{-3}-10^{3}$Hz GWs during their inspiral, merger and ring-down phases, whose low-frequency end would be detected by future space-based interferometers~\cite{LISA:2017pwj,Hu:2017mde,TianQin:2015yph} and whose high-frequency end can be detected by ground-based interferometers~\cite{KAGRA:2021vkt,LIGOScientific:2020ibl,LIGOScientific:2018mvr}; GWs at frequencies higher than $10$kHz are proposed to be sourced by some phenomenons involving beyond the Standard Model physics, such as preheating after inflation~\cite{Khlebnikov:1997di,Easther:2006gt,Garcia-Bellido:2007fiu} and phase transitions at high energies in the early Universe~\cite{Grojean:2006bp,Hindmarsh:2015qta,Hindmarsh:2013xza,Kosowsky:1992rz}, which prompt many new detector concepts in the laboratory~\cite{Aggarwal:2020olq}. What interesting astrophysical objects or cosmological events can source GWs in the $\lesssim10^{-8}$Hz frequency range?

As a promising candidate for dark matter~\cite{Rubin:1982kyu,Davis:1985rj,Clowe:2006eq}, the ultralight scalar field with spin-$0$, extraordinarily light mass ($m\sim10^{-22}\rm{eV}/c^2$) and de Broglie wavelength comparable to a few kpc, namely fuzzy dark matter (FDM)~\cite{Hu:2000ke}, can form an equilibrium configuration with size smaller than its de Broglie wavelength, the so-called FDM soliton~\cite{Guzman:2004wj,Davies:2019wgi}. The subsequent evolution of FDM solitons are usually simulated numerically, including the perturbation, the interference/collision and the tidal disruption/deformation of FDM solitons~\cite{Guzman:2004wj,Paredes:2015wga,Edwards:2018ccc,Bernal:2006ci,Munive-Villa:2022nsr,Koo:2025qac,Schwabe:2016rze}, according to the coupled Schrödinger–Poisson (SP) system of equations 
\begin{equation}
\label{eq:sp0} 
\begin{cases}
\begin{aligned}
&i\hbar\frac{\partial \Psi}{\partial t}=\left(-\frac{\hbar^2}{2m}\nabla^2+m\Phi\right)\Psi, \\
&\nabla^2 \Phi=4\pi G m|\Psi|^2, 
\end{aligned}
\end{cases}
\end{equation}
where FDM is described by the \Mov{complex} wavefunction $\Psi$, $m$ is the mass of FDM and the gravitational potential $\Phi$ is sourced by the FDM density $\rho=m
|\Psi|^2$. 
Because of the large size of the FDM solitions and the weak self-gravitational potential, it is hard to form gravitational bound systems between FDM solitons, such as binary FDM solitons. Even though without the normal inspiral, merger and ring-down phases of a compact binary system, 
the other spherically or rotationally asymmetric evolution of FDM solitons should also emit GWs.
As shown in this paper, for example, since the size of the FDM soliton is typically on the order of kpc, the frequency of GWs emitted from collision of FDM solitons is typically on the order of c/kpc$\sim10^{-11}$Hz. The mass $m$ of the FDM particle or the velocity $v$ and mass $M$ of the FDM solitons can further be fine-tuned to adjust the frequency of such GWs. In other words, future detection of such GWs can constrain the property of the FDM particle and the FDM solitons. 

However, gravitational radiation carries away the energy and momentum of isolated systems. Therefore, in principle, the SP system (Eq.~(\ref{eq:sp0})) should be enlarged by including the wave equation for GWs $h_{\mu\nu}$
\begin{equation}
\label{eq:wave}
\square h_{\mu\nu}=-\frac{16\pi G}{c^4} T_{\mu\nu},
\end{equation}
where $T_{\mu\nu}$ is the energy-momentum tensor of FDM solitons and we have assumed that the linearized theory is valid.
\Ke{The compactness of solitons calculated at the end of Section~\ref{sec:soliton2} not only validates the assumption of the linearized theory, but also is consistent with the common finding that} the SP system is nothing but the weak field limit of its general relativistic
counterpart, the Einstein–Klein–Gordon (EKG) system~\cite{Kaup:1968zz,Ruffini:1969qy,Ma:2023vfa}. Furthermore, the GW back reaction on FDM solitons will modify Eq.~(\ref{eq:sp0}). As the system's energy is carried away by GWs, the FDM solitons would settle down gradually and eventually. However, if the gravitational radiation is not efficient, the GW back reaction on the evolution of FDM solitons can be neglected approximately, which is done in this paper.

This paper is organized as follows.
In Section~\ref{sec:soliton1}, we turn to the shooting method to solve the SP system (Eq.~(\ref{eq:sp0})) and obtain the isolated FDM solitons.
In Section~\ref{sec:soliton2}, we numerically simulate the evolution of FDM solitons with $\mathtt{PyUltraLight}$~\cite{Edwards:2018ccc}, the collision of FDM solitons in particular.
In Section~\ref{sec:gw}, we calculate the waveform of GWs emitted from collision of FDM solitons.
Finally, a brief summary and discussions are included in Section~\ref{sec:sd}.

\section{Fuzzy Dark Matter Solitons}
\label{sec:soliton1}
\begin{figure}[]
\begin{center}
\subfloat{\includegraphics[width=0.5\textwidth]{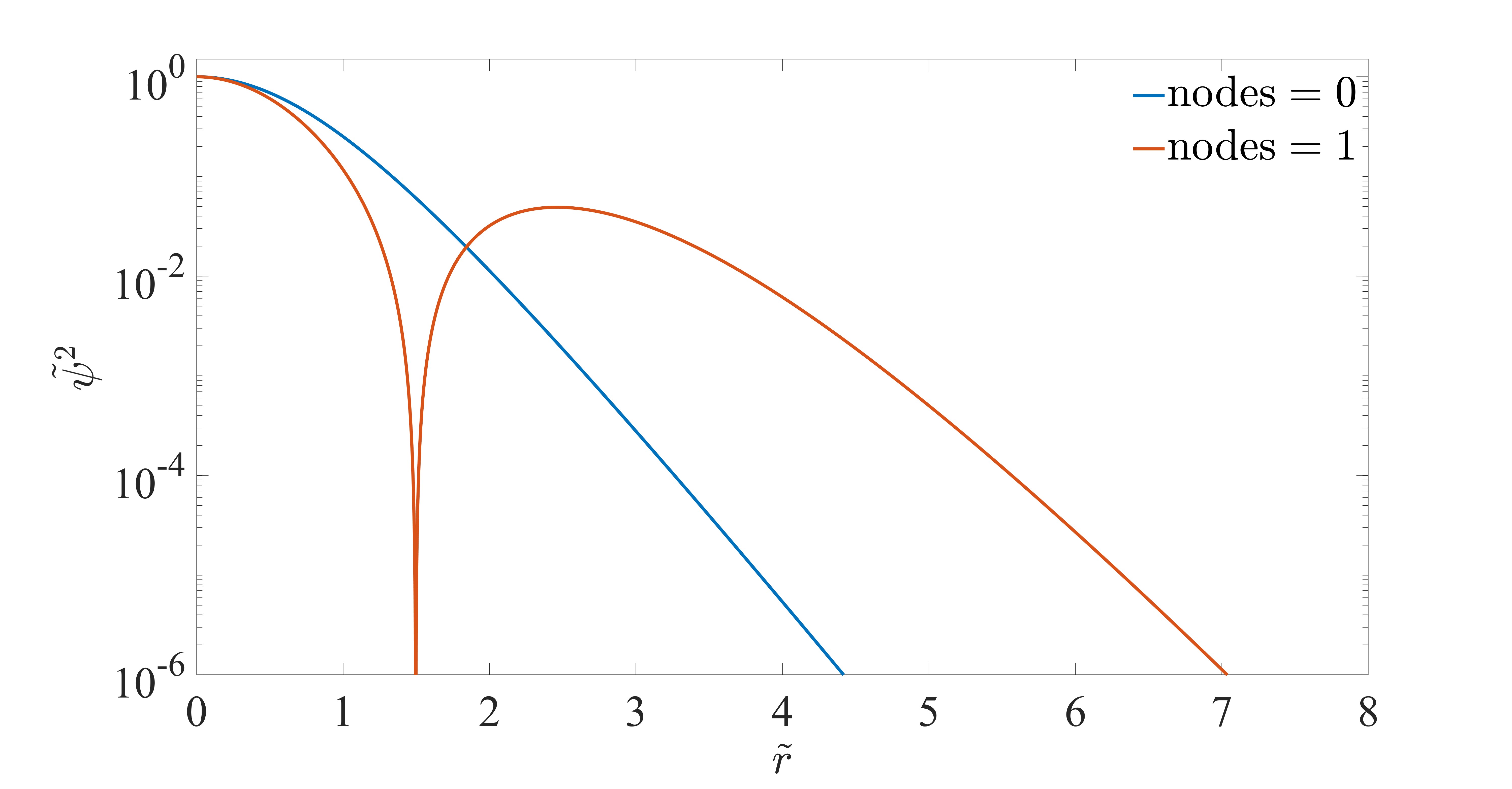}}
\end{center}
\captionsetup{justification=raggedright}
\caption{Density profiles of the ground state (blue curve) and the first excited state (orange curve).}
\label{fig:pf}
\end{figure}

Before the numerical simulation of the evolution of FDM solitons, we should first build up the profile of an isolated FDM soliton.
Since an isolated FDM soliton features spherical symmetry, the ansatz of $\Psi(r, t) =e^{i\gamma t}\psi(r)$ 
\Tan{works. Here $|\Psi|^2$ can be interpreted as the FDM particle number density according to Poisson equation in Eq.~(\ref{eq:sp0}). As a result, }the FDM soliton density is $\rho(r)=m|\Psi|^2=m\psi^2(r)$ and the FDM soliton mass is $M=\int_0^\infty4\pi r^2\rho(r)dr$. 
After defining a number of dimensionless variables as
\begin{align}
\label{eq:dless1}
\nonumber
&\tilde{r}\equiv\frac{mc}{\hbar}r,\\\nonumber
&\tilde{t}\equiv\frac{mc^2}{\hbar}t,\\\nonumber
&\tilde{\psi}\equiv\frac{\hbar\sqrt{ G}}{c^2\sqrt{m}} \psi,\\\nonumber
&\tilde{\Phi}\equiv\frac{1}{c^2}\Phi,\\\nonumber
&\tilde{\gamma}\equiv\frac{\hbar}{mc^2}\gamma,\\
&\tilde{M} \equiv \frac{Gm}{\hbar c}M,
\end{align}
the dimensionless spatial part of Eq.~(\ref{eq:sp0}) is
\begin{equation}
\label{eq:sp1} 
\begin{cases}
\begin{aligned}
&\frac{\partial^2 (\tilde{r}\tilde{\psi})}{\partial \tilde{r}^2}=2\tilde{r}\left(\tilde{\Phi}+\tilde{\gamma}\right)\tilde{\psi},\\
&\frac{\partial^2 (\tilde{r}\tilde{\Phi})}{\partial \tilde{r}^2}=4\pi\tilde{r}\tilde{\psi}^2.
\end{aligned}
\end{cases}
\end{equation}
Fulfilling arbitrary normalization $\tilde{\psi}(\tilde{r}=0)=1$ and boundary conditions $\tilde{\Phi}(\tilde{r}=\infty)=0$, $\tilde{\psi}( \tilde{r}=\infty)=0$,
$\frac{\partial\tilde{\psi}}{\partial\tilde{r}}|_{\tilde{r}=0}=0$ and $\frac{\partial \tilde{\Phi}}{\partial\tilde{r}}|_{\tilde{r}=0}=0$ and adjusting the quantized eigenvalue $\tilde{\Phi}(\tilde{r}=0) +\tilde{\gamma}$, we can calculate the equilibrium configurations from Eq.~(\ref{eq:sp0}) by the shooting method. Only the solution from the smallest $\tilde{\gamma}$ is stable and the ground state. 
We also obtain the first excited state.
In Tab.~\ref{tb:pf}, we list the eigenvalues $\tilde{\gamma}$ and soliton masses $\tilde{M}$ of the ground state and the first excited state. In Fig.~\ref{fig:pf}, the corresponding soliton profiles are plotted.   
Since the SP system (Eq.~(\ref{eq:sp1})) has the scaling symmetry
\begin{align}
\label{eq:sc}
\nonumber
&\tilde{r}  \longrightarrow  \lambda^{-1/2} \tilde{r},\\\nonumber
&\tilde{\psi}  \longrightarrow  \lambda \tilde{\psi}, \\\nonumber
&\tilde{\Phi}  \longrightarrow  \lambda \tilde{\Phi}, \\\nonumber
&\tilde{\gamma}  \longrightarrow  \lambda \tilde{\gamma}, \\
&\tilde{M} \longrightarrow  \lambda^{1/2} \tilde{M},
\end{align}
we can adjust the mass of FDM solitons with \Mov{an arbitrary scaling constant} $\lambda$ to set up different initial conditions for collision.
\begin{table}[htbp] 
\renewcommand\arraystretch{1.5}
\captionsetup{justification=raggedright}
\caption{Eigenvalues $\tilde{\gamma}$ and masses $\tilde{M}$ of the ground state and the first excited state, \Tan{where the number of nodes is the number of zero-crossings of the profile $\tilde{\psi}(\tilde{r})$, as hinted in Fig.~\ref{fig:pf}.}} 
\label{tb:pf}
\begin{tabular}{ | c | c| c | } 
\hline Nodes & $\tilde{\gamma}$ & $\tilde{M}$\\ 
\hline 
0 & \ 2.45 & 3.88\\ 
1 & \ 2.30 & 8.64\\ 
\hline
\end{tabular}
\end{table}

Summarizing Eq.~(\ref{eq:dless1}), we can introduce the length, time and mass scales as follows
\begin{align}
\label{eq:unit}
\nonumber
&\mathcal{L}\equiv\frac{\hbar}{mc}\approx6.39\times 10^{-5}\left(\frac{10^{-22}\rm{eV}/c^2}{m}\right)\rm{kpc},\\\nonumber
&\mathcal{T}\equiv\frac{\hbar}{mc^2}\approx6.57\times 10^{6}\left(\frac{10^{-22}\rm{eV}/c^2}{m}\right)\rm{s},\\
&\mathcal{M}\equiv\frac{\hbar c}{Gm}\approx1.34\times 10^{12}\left(\frac{10^{-22}\rm{eV}/c^2}{m}\right)\rm{M}_{\odot}.
\end{align}
Since, for example, the mass of an isolated soliton $M$ can be predicted from the halo mass $M_{\rm{halo}}$ according to the soliton-halo mass relation~\cite{Schive:2014hza}
\begin{align} 
\label{eq:halo0}
\nonumber
M &\approx 1.25\times 10^{9}\left(\frac{M_{\rm{halo}}}{10^{12}\rm{M}_{\odot}}\right)^{1/3} \left(\frac{m}{10^{-22}{\rm{eV}}/c^2}\right)^{-1}\rm{M}_{\odot}\\
&\approx 9.33\times 10^{-4}\left(\frac{M_{\rm{halo}}}{10^{12}\rm{M}_{\odot}}\right)^{1/3}\mathcal{M},
\end{align}
for the ground state, we have
\begin{equation}
\label{eq:halo1}
\lambda\approx5.78\times10^{-8}\left(\frac{M_{\rm{halo}}}{10^{12}\rm{M}_{\odot}}\right)^{2/3};
\end{equation}
for the first excited state, we have
\begin{equation}
\label{eq:halo2}
\lambda\approx1.17\times10^{-8}\left(\frac{M_{\rm{halo}}}{10^{12}\rm{M}_{\odot}}\right)^{2/3}. 
\end{equation}
Then, for a more general system of FDM solitons, we can obtain a number of dimensionless variables including the spatial position, time and the wavefunction, gravitational potential, eigenvalue, mass, density, energy~\cite{Edwards:2018ccc} and velocity of FDM solitons respectively as 
\begin{align}
\label{eq:dless2}
\nonumber
&\tilde{\vec{x}}=\frac{\vec{x}}{\mathcal{L}},\\\nonumber
&\tilde{t}=\frac{t}{\mathcal{T}},\\\nonumber
&\tilde{\Psi}=\mathcal{T}\sqrt{mG}\Psi,\\\nonumber
&\tilde{\Phi}=\frac{m\mathcal{T}}{\hbar}\Phi,\\\nonumber
&\tilde{\gamma}=\mathcal{T}\gamma,\\\nonumber
&\tilde{M}=\frac{M}{\mathcal{M}}=\int d^3\tilde{x}|\tilde{\Psi}|^2,\\\nonumber
&\tilde{\rho}=\frac{\rho}{\mathcal{ML}^{-3}}=|\tilde{\Psi}|^2,\\\nonumber
&\tilde{E}=\frac{E}{\mathcal{ML}^{2}\mathcal{T}^{-2}}=\int d^{3}x\left(\frac{1}{2}\tilde{\Phi}|\tilde{\Psi}|^{2}-\frac{1}{2}\tilde{\Psi}^{*}\nabla^{2}\tilde{\Psi}\right),\\
&\tilde{\vec{v}}=\frac{\vec{v}}{\mathcal{L}\mathcal{T}^{-1}}.
\end{align}
Finally, the dimensionless version of Eq.~(\ref{eq:sp0}) without any assumption of spatial symmetry is
\begin{equation}
\label{eq:sp2}
\begin{cases}
\begin{aligned}
&i\frac{\partial \Psi(\vec{x},t)}{\partial t}=\left(-\frac{1}{2}\nabla^{2}+\Phi(\vec{x},t)\right)\Psi(\vec{x},t), \\
&\nabla^2\Phi(\vec{x},t)=4\pi|\Psi(\vec{x},t)|^2,
\end{aligned}
\end{cases}
\end{equation}
where we have dropped the tildes for notational convenience until Section~\ref{sec:gw}.

\section{Collision of Fuzzy Dark Matter Solitons}
\label{sec:soliton2}
In this paper, we numerically simulate the evolution of FDM solitons with $\mathtt{PyUltraLight}$~\cite{Edwards:2018ccc}, the collision of FDM solitons in particular. $\mathtt{PyUltraLight}$ is designed to solve the SP system \Mov{in a cubic domain} under periodic boundary conditions. Then, Eq.~(\ref{eq:sp2}) should be rewrite as 
\begin{equation}
\label{eq:sp3} 
\begin{cases}
\begin{aligned}
&i\frac{\partial \Psi(\vec{x},t)}{\partial t}=\left(-\frac{1}{2}\nabla^2+\Phi(\vec{x},t)\right)\Psi(\vec{x},t), \\
&\nabla^2 \Phi(\vec{x},t)=4\pi (|\Psi(\vec{x},t)|^2-\langle|\Psi(\vec{x},t)|^2\rangle), 
\end{aligned}
\end{cases}
\end{equation}
where $\langle|\Psi(\vec{x},t)|^2\rangle$ is the average density over the simulation box. After a sufficiently small timestep $\Delta t$, $\mathtt{PyUltraLight}$ evolves $\Psi(\vec{x}, t)$ and $\Phi(\vec{x}, t)$ in Eq.~(\ref{eq:sp3}) as
\begin{equation}
\label{eq:sp4}
\begin{cases}
\begin{aligned}
\Psi(\vec{x},t+\Delta t/2) &= \mathcal{F}^{-1}\exp\left[-\frac{i\Delta t}{2}k^{2}\right]\mathcal{F}\\
&\times\exp\left[-\frac{i\Delta t}{2}\Phi(\vec{x},t)\right] \Psi(\vec{x},t),\\
\Phi(\vec{x},t+\Delta t)&=\mathcal{F}^{-1}\left(-\frac{1}{k^2}\right)\mathcal{F}4\pi|\Psi(\vec{x},t+\Delta t/2)|^2,\\
\Psi(\vec{x},t+\Delta t) &= \exp\left[-\frac{i\Delta t}{2}\Phi(\vec{x},t+\Delta t)\right] \Psi(\vec{x},t+\Delta t/2),
\end{aligned}
\end{cases}
\end{equation}
where the order of operations runs from right to left, $\mathcal{F}$ and $\mathcal{F}^{-1}$ denote the discrete Fourier transform and its inverse, $k$ is the wavenumber in the Fourier domain and the global average density disappears.
\Mov{Here, $\mathtt{PyUltraLight}$~\cite{Edwards:2018ccc} turns to the finite difference method to discretise $\Phi(\vec{x})$ and $\Psi(\vec{x})$ but the spectral method to deal with the operator $\nabla^2$.}
For the collision of two FDM solitons, the initial total wavefunction is
\begin{align}
\nonumber
\Psi(\vec{x},t_0)=\sum_{n=1}^2&\left[\lambda_{n}\psi(\sqrt{\lambda_{n}}|\vec{x}-\vec{x}_{n}-\vec{v}_{n}t_0|)~\times\right.\\
&\left.e^{i(\lambda_{n}\gamma_{n} t_0+\vec{v}_{n}\cdot(\vec{x}-\vec{x}_{n})-\frac{1}{2}|\vec{v}_{n}|^{2}t_0+\delta_n)}\right],
\end{align}
where $\lambda_n$ rescales the initial mass and size of FDM soliton according to Eq.~(\ref{eq:sc}), $\gamma_n$ is the soliton eigenvalue (or $\lambda_n\gamma_n$ is its rescaled counterpart), $\vec{x}_{n}$ is the initial central position of the FDM solition, $\vec{v}_{n}$ is the FDM soliton velocity, $\delta_n$ is the initial phase and $t_0$ is the start time. Given $t_0=0$, the simulation resolution \Ke{$N^3=256^3$ (or $N^3=768^3$)} and the length of simulation box \Ke{$l=4\times10^4$ (or $l=12\times10^4$)}, we can study the dynamics of the FDM soliton collisions. In Tab.~\ref{tb:set}, we list the setup for $3$ head-on collisions, namely C$1$, C$2$ and C$3$\footnote{The periodic boundary conditions do play an important role during the simulation. The smaller domain size, the more non-negligible images of each soliton due to the periodic boundary conditions. Unless the domain size is large enough, the potential of images should be subtracted from $\Phi$ in Eq.~(\ref{eq:sp4}).
For example, we find that one simulation with $124$ nearby images been subtracted for each soliton, where the initial separation between two solitons is $2\times10^4$ in a cubic domain with length $4\times10^4$, converges to another simulation without images-subtraction, where the initial separation between two solitons is $2\times10^4$ in a cubic domain with length $12\times10^4$. 
Since the density profile of the first excited state is more complicated than that of the ground state, as shown in Fig.~\ref{fig:pf}, images-subtraction is not suitable for the soliton at the first excited state. Finally, we turn to $l=4\times10^4$ with images-subtraction for collisions C$1$ and C$2$, but $l=12\times10^4$ without images-subtraction for collision C$3$. For all simulations, we fix the spatial grid spacing as $\frac{4\times10^4}{256}=\frac{12\times10^4}{768}$.\\
Without the GW back reaction on the evolution of FDM solitons, our simulations of the dimensionless version of SP system (Eq.~\ref{eq:sp3}) are exactly the same as the simulations of Eqs.~(2.13) and (2.14) in $\mathtt{PyUltraLight}$~\cite{Edwards:2018ccc}. Since, in section 5 of~\cite{Edwards:2018ccc}, a multifaceted approach to convergence testing has been adopted, here we just directly turn to their conclusion about the convergence analysis that ``We also see that as the spatial resolution is decreased from $320^3$ to $64^3$, the lowest resolution performs poorly, but there is good convergence at resolutions of $192^3$ and above.". 
\Tan{For example, the total energy $E(t)$ converges at an apparent order of approximately $3.1$ according to Richardson Extrapolation and the fractional error of $E(t)$ between resolutions of $192^3$ and $256^3$ is $\frac{ E_{192}(t)- E_{256}(t)}{E_{256}(t)}\sim 2\times 10^{-6}$.}}
\begin{table}[htbp] 
\renewcommand\arraystretch{1.5}
\captionsetup{justification=raggedright}
\caption{Setup for $3$ head-on collisions, where $\gamma_n$ decides the initial density profile according to Tab.~\ref{tb:pf} and Fig.~\ref{fig:pf} and $\lambda_n$ is related to its corresponding $M_{\rm{halo}}$ as Eq.~(\ref{eq:halo1}) or Eq.~(\ref{eq:halo2}).} 
\label{tb:set}
\begin{tabular}{ |c | c  c  c  c  c| } 
\hline 
Collision & $\lambda_1$ & $\gamma_1$ & $\delta_1$ & $\vec{x}_1$ & $\vec{v}_1$\\ 
\hline 
C$1$ & $1.66\times 10^{-6}$ & $2.45$ & $0$ & $\left(\Ke{10^4},0,0\right)$ & $\left(10^{-4},0,0\right)$\\
C$2$ & $1.66\times 10^{-6}$ & $2.45$ & $0$ & $\left(\Ke{10^4},0,0\right)$ & $\left(10^{-4},0,0\right)$\\
C$3$ & $1.66\times 10^{-6}$ & $2.45$ & $0$ & $\left(\Ke{10^4},0,0\right)$ & $\left(10^{-4},0,0\right)$\\
\hline 
Collision & $\lambda_2$ & $\gamma_2$ & $\delta_2$ & $\vec{x}_2$ & $\vec{v}_2$\\ 
\hline 
C$1$ & $1.66\times 10^{-6}$ & $2.45$ & $0$ & $\left(\Ke{3\times10^4},0,0\right)$ & $\left(-10^{-4},0,0\right)$\\
C$2$ & $4.15\times 10^{-7}$ & $2.45$ & $0$ & $\left(\Ke{3\times10^4},0,0\right)$ & $\left(-10^{-4},0,0\right)$\\
C$3$ & $1.34\times 10^{-6}$ & $2.30$ & $0$ & $\left(\Ke{3\times10^4},0,0\right)$ & $\left(0,0,0\right)$\\
\hline
\end{tabular}
\end{table}

\begin{figure*}[]
\begin{center}
\subfloat{\includegraphics[height=0.35\textheight]{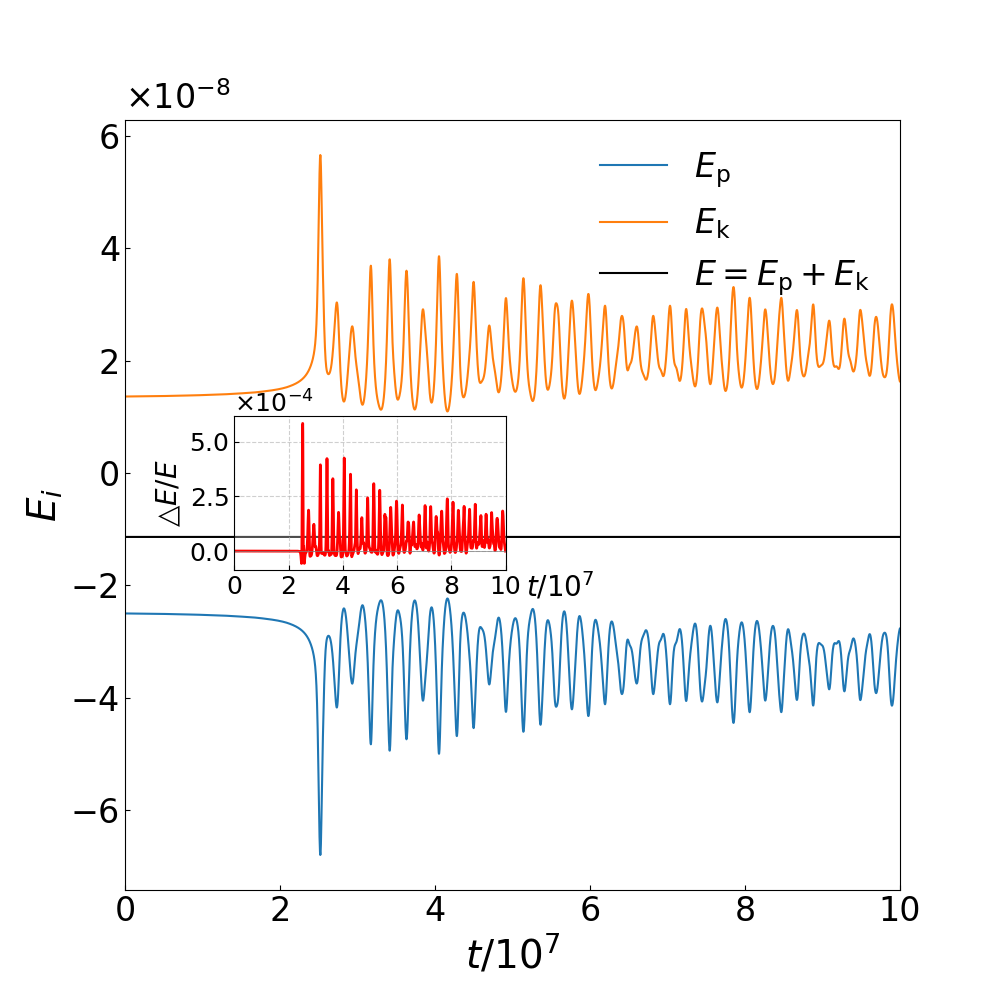}}
\subfloat{\includegraphics[height=0.35\textheight]{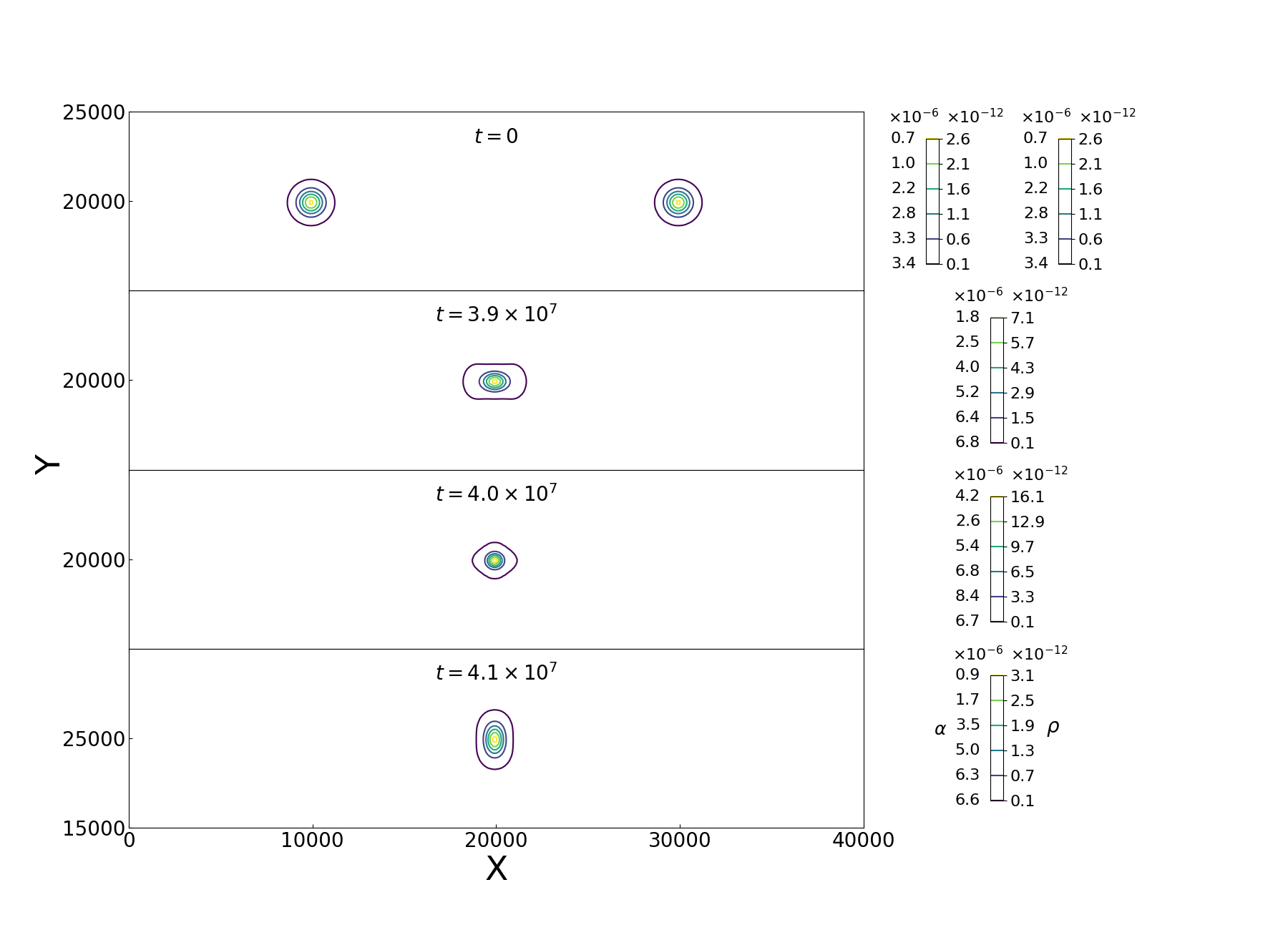}}

\end{center}
\captionsetup{justification=raggedright}
\caption{\Tan{Simulation of C1}. LEFT: Energy transfer between the kinetic energy $E_{\rm{k}}$ and the gravitational potential energy $E_{\rm{p}}$\Mov{, where energy conservation is guaranteed by an almost constant of $E=E_{\rm{p}}+E_{\rm{k}}$ with the fractional error about \Ke{$\sim10^{-4}$}}. RIGHT: Snapshots of the evolution of FDM solitons, where \Tan{each density contour not only represents a line of constant $\rho$, but also corresponds to a density isosurface that encloses a volume with a derived compactness parameter $\alpha$. 
The values of density $\rho$ and compactness $\alpha$ for density contours are indicated by the side bars: for each side color bar, the left number is $\alpha$ and the right value is $\rho$; the two side color bars in the top panel are for the left and right FDM soliton, respectively.} For the initial configuration \Tan{at $t=0$}, the largest density of each soliton is $\lambda_n^2$ of C1 listed in Tab.~\ref{tb:set}.}
\label{fig:C1}
\end{figure*}

\begin{figure*}[]
\begin{center}
\subfloat{\includegraphics[height=0.35\textheight]{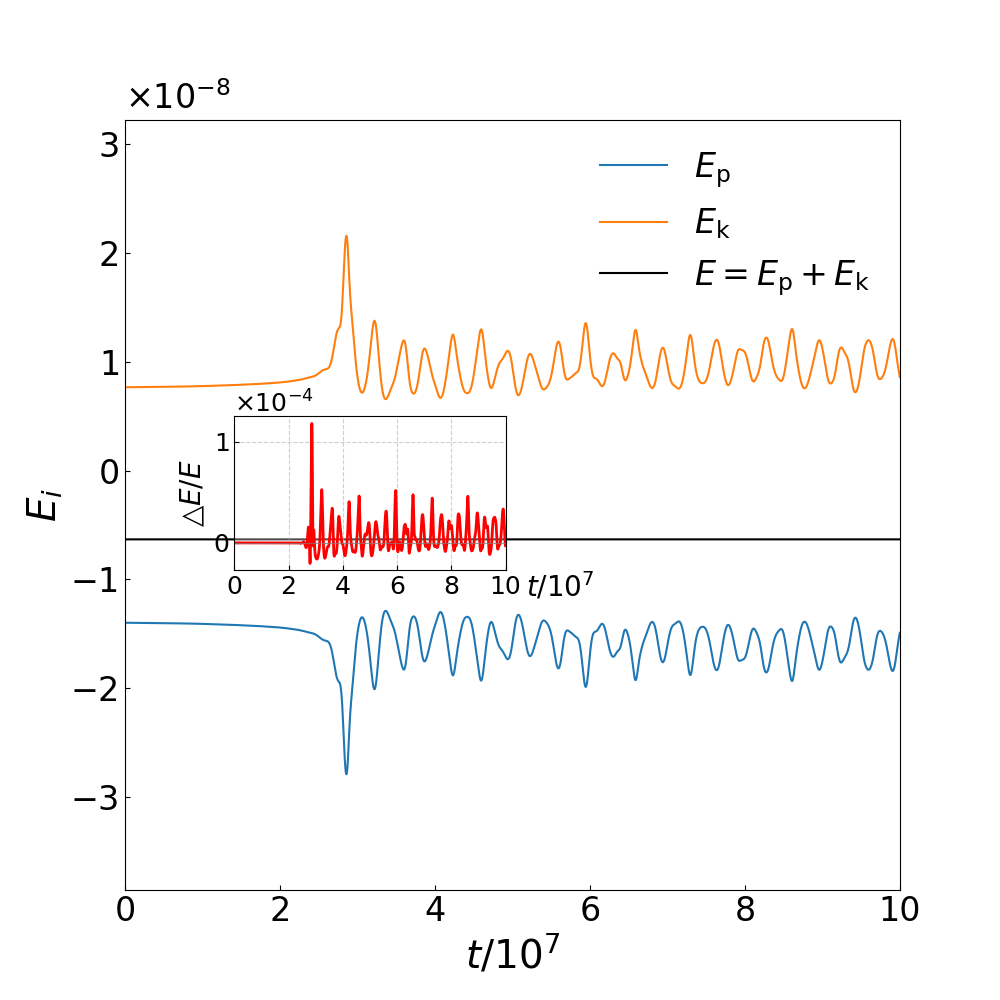}}
\subfloat{\includegraphics[height=0.35\textheight]{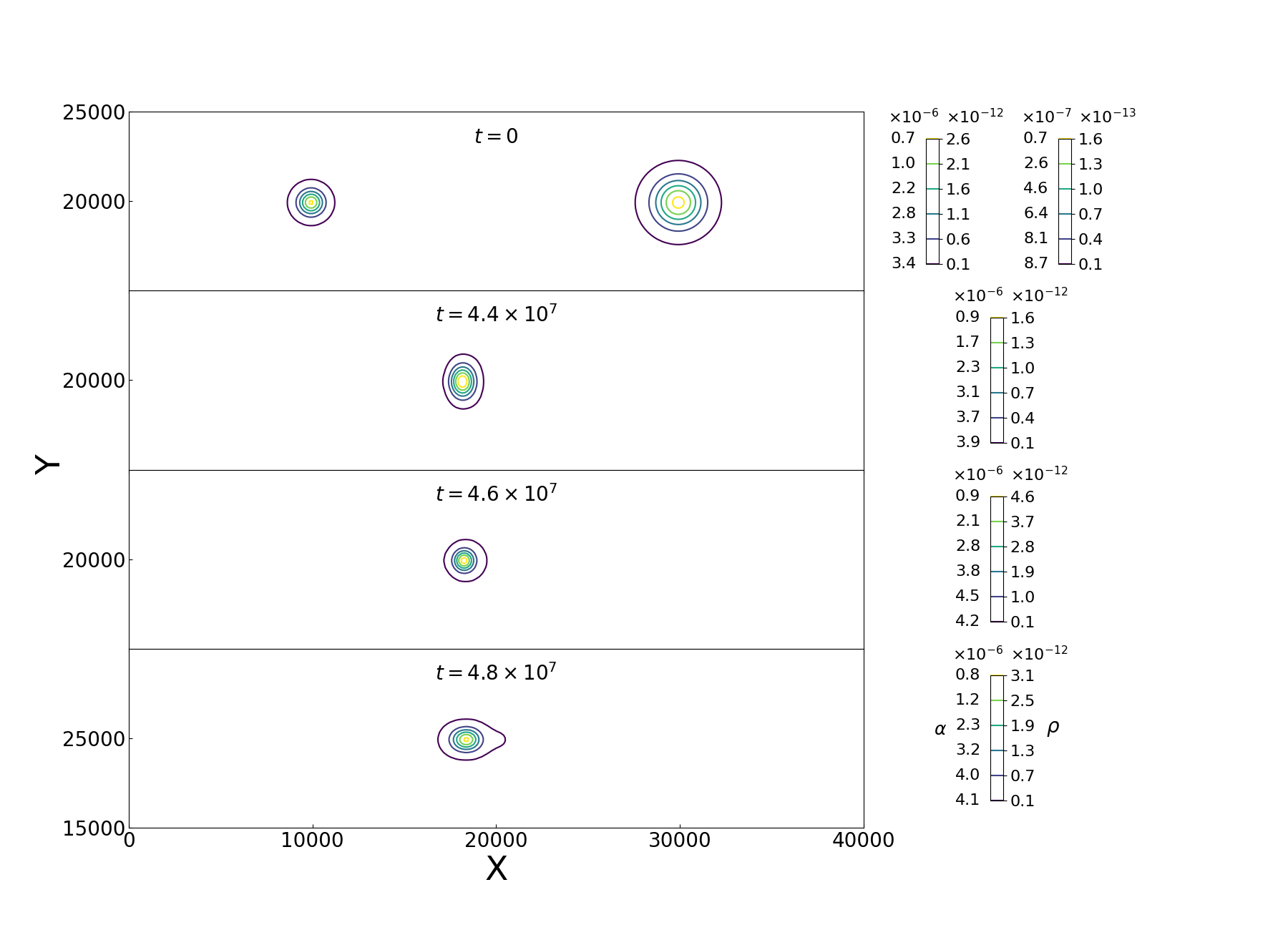}}
\end{center}
\captionsetup{justification=raggedright}
\caption{\Tan{Same as Fig.~\ref{fig:C1} but for C2.} 
}
\label{fig:C2}
\end{figure*}

\begin{figure*}[]
\begin{center}
\subfloat{\includegraphics[height=0.35\textheight]{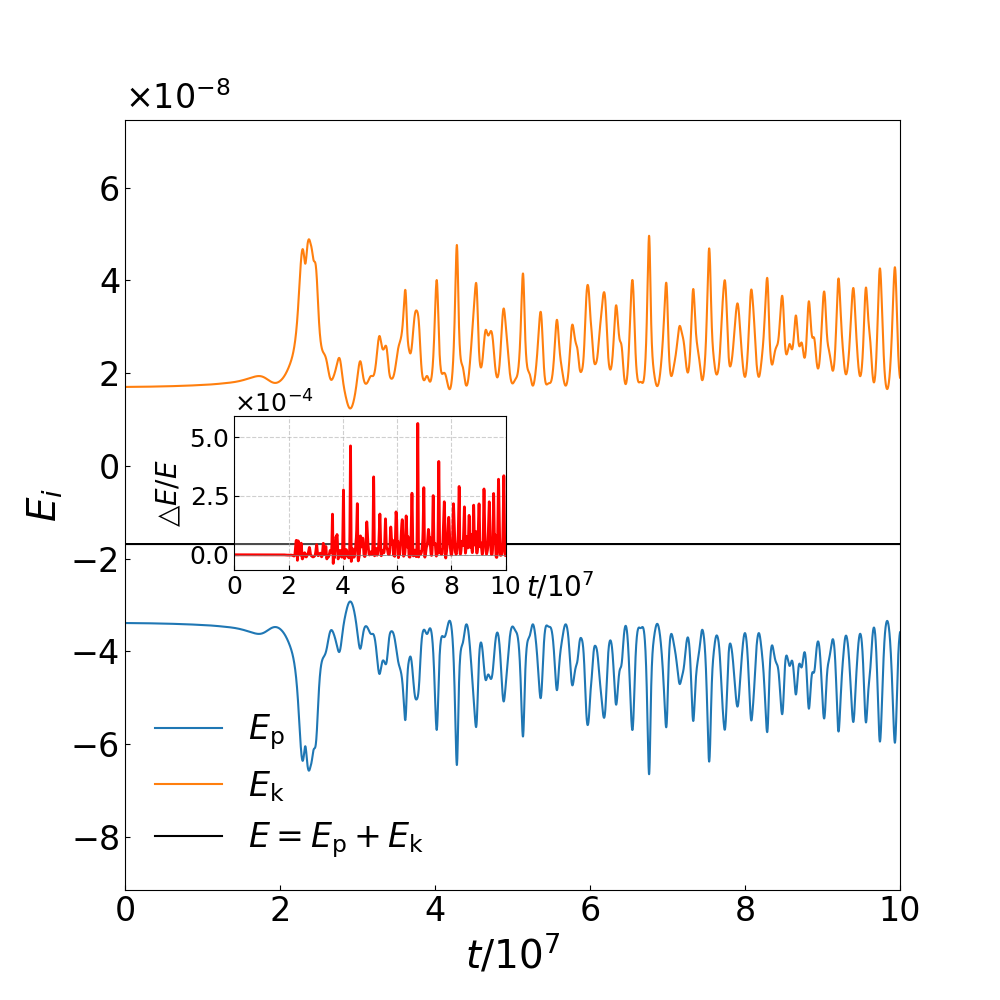}}
\subfloat{\includegraphics[height=0.35\textheight]{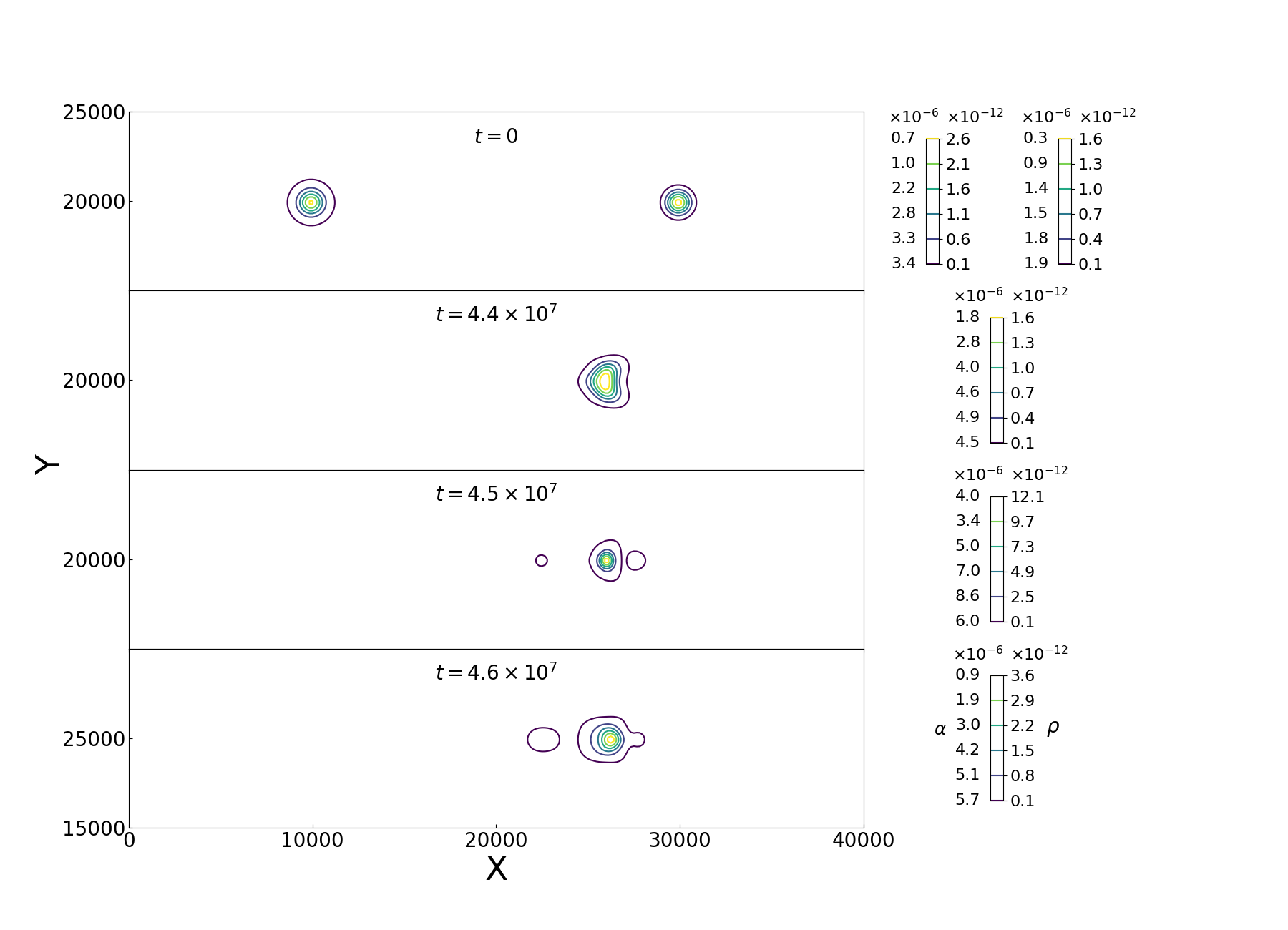}}
\end{center}
\captionsetup{justification=raggedright}
\caption{\Tan{Same as Fig.~\ref{fig:C1} but for C3.} 
}
\label{fig:C3}
\end{figure*}

During simulations, energy conservation should be guaranteed. As shown in Eq.~(\ref{eq:dless2}), the total energy \Mov{is a constant and} can be decomposed into the kinetic energy $E_{\rm{k}}$ and the gravitational potential energy $E_{\rm{p}}$ when the GW back reaction is ignored. In the left subplots of Fig.~\ref{fig:C1}, Fig.~\ref{fig:C2} and Fig.~\ref{fig:C3}, we find that energy conservation is satisfied.
In the right subplots of Fig.~\ref{fig:C1}, Fig.~\ref{fig:C2} and Fig.~\ref{fig:C3}, we show the evolution of FDM solitons for head-on collision. We find that $|E_{\rm{p}}|$ is large enough to merge the initial two FDM solitons with each other after collisions; the energy transfer between $E_{\rm{k}}$ and $E_{\rm{p}}$ causes the final FDM soliton to oscillate irregularly with an adimensional frequency\footnote{\Ke{Here the frequency is an adimensional one. Its physical implications are only clear when the FDM particle mass $m$ (or $\mathcal{T}$) is given in Eq.~(\ref{eq:unit}). For $m=10^{-22}\rm{eV}/c^2$, for example, the final FDM soliton oscillates at $\sim10^{-14}$Hz, which is commensurate with the finding of~\cite{Li:2020ryg} for gravitational collapse and collision.}} of \Ke{$\sim10^{-7}$}; the location of merger depends on system's mass-ratio. Especially for C3, the mass-ratio between one FDM soliton with the density profile as the ground state and another with the density profile as the first excited state is $0.5$ but the more massive one without initial velocity is attracted away from its initial position, as shown in the right subplot of Fig.~\ref{fig:C3}. It means that the FDM soliton with the density profile as the first excited state is not stable and is dismembered before collision.
\Mov{Also in these right subplots, the side bars indicate the values of density $\rho$ and compactness $\alpha$ for density contours. Here, for a certain density \Tan{isosurface}, $\alpha$ is derived from the ratio of the total mass inside it to its minimum radius. We find that, for every density contour of each snapshot in Figs.~\ref{fig:C1}-\ref{fig:C3}, $\alpha$ is comparable to the compactness of the Sun $2.12\times10^{-6}$. Therefore, our assumption of the linearized approach (Eq.~(\ref{eq:wave})) is reasonable.}

\section{Gravitational Waves from Collision}
\label{sec:gw}

\begin{figure*}[]
\begin{center}
\subfloat{\includegraphics[width=1.0\textwidth]{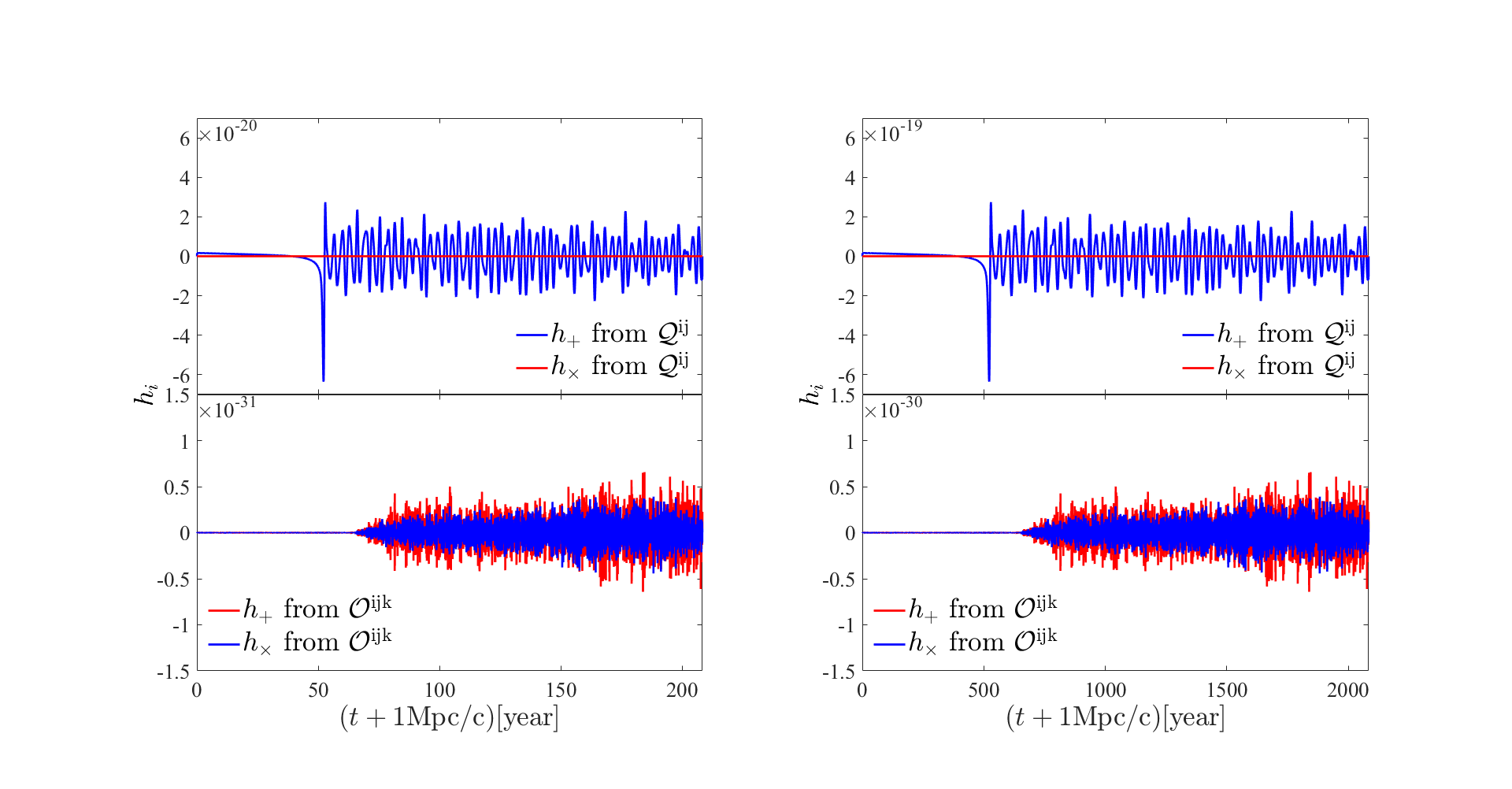}}
\end{center}
\captionsetup{justification=raggedright}
\caption{GWs with $h_{+}$ polarization (blue curves) and $h_{\times}$ polarization (red lines) from C1, where the distance from source is $1$Mpc and the FDM mass is $m=10^{-17}\rm{eV}/c^2$ (left) or $m=10^{-18}\rm{eV}/c^2$ (right).}
\label{fig:gw1}
\end{figure*}

\begin{figure*}[]
\begin{center}
\subfloat{\includegraphics[width=1.0\textwidth]{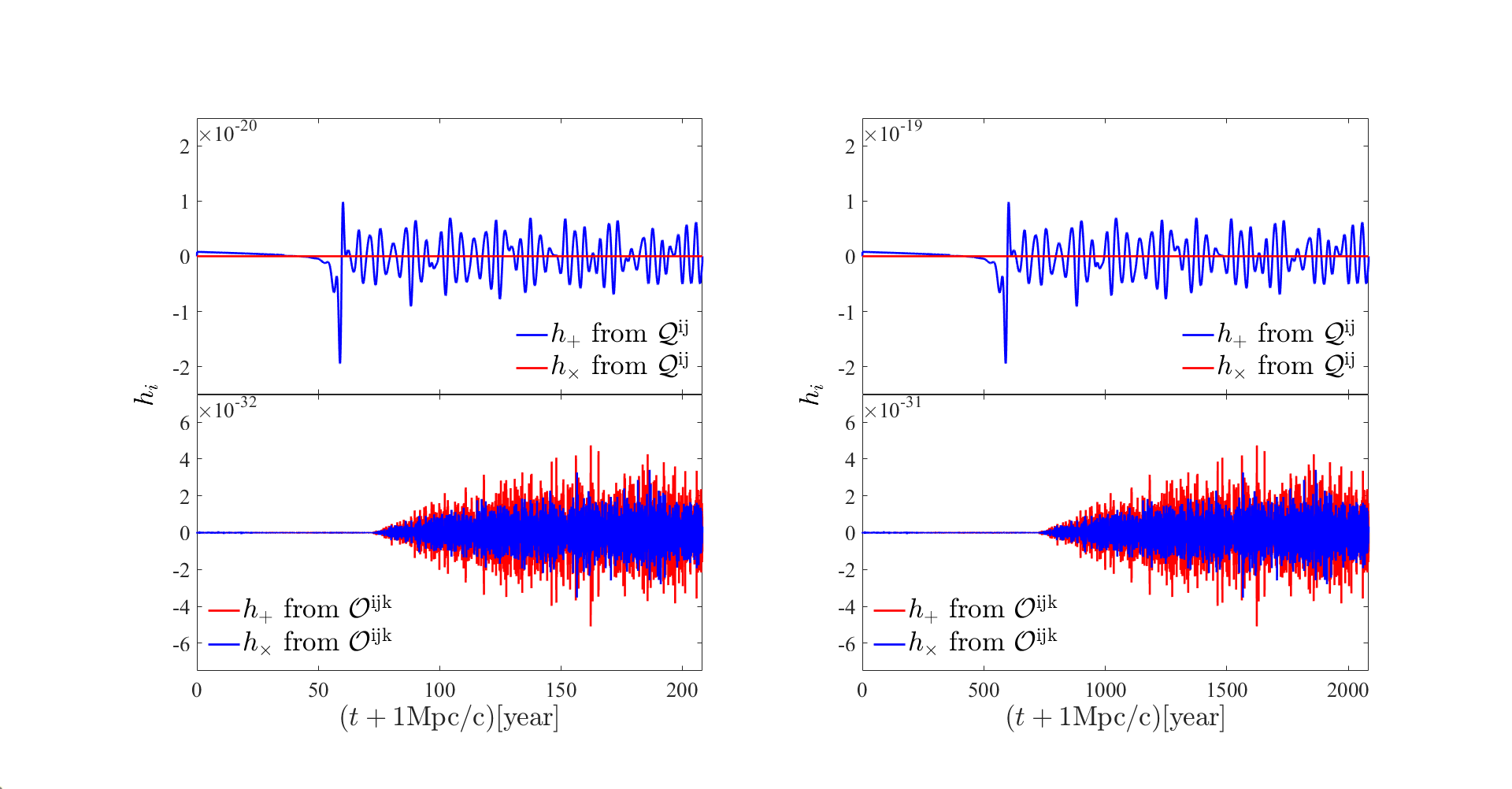}}
\end{center}
\captionsetup{justification=raggedright}
\caption{GWs with $h_{+}$ polarization (blue curves) and $h_{\times}$ polarization (red lines) from C2, where the distance from source is $1$Mpc and the FDM mass is $m=10^{-17}\rm{eV}/c^2$ (left) or $m=10^{-18}\rm{eV}/c^2$ (right).}
\label{fig:gw2}
\end{figure*}

Similarly to the GWs from head-on collisions of two Proca stars~\cite{CalderonBustillo:2020fyi}, GWs are also emitted from collisions of FDM solitons. Since the SP system itself is the weak field limit of the EKG system, we suppose that the linearized theory is valid for calculation of GWs from collisions of FDM solitons. Solving the wave equation for GWs (Eq.~(\ref{eq:wave})) on transverse-traceless gauge, we obtain the multipole expansion of GWs~\cite{Blanchet:1995fg,Roskill:2023bmd}
\begin{align}
\nonumber
h_{ij}^{TT}(\vec{x},t)&=\frac{1}{r}\frac{2G}{c^4}\Lambda_{ij,kl}\ddot{\mathcal{Q}}^{kl}\\\nonumber
&\quad +\frac{1}{r}\frac{2G}{c^5}\Lambda_{ij,kl}\left[\frac{1}{3}n_{m}\dddot{\mathcal{O}}^{klm}+\frac{2}{3}n_{m}\ddot{\mathcal{Z}}^{klm}\right]\\
&\quad +...,
\end{align}
where $\Lambda_{ij,kl}$ is the projection operator orthogonal to the normal vector $n^i$
\begin{align}\nonumber
\nonumber
\Lambda_{ij,kl}&=P_{ik}P_{jl}-\frac{1}{2}P_{ij}P_{kl},\\
P_{ij}&=\delta_{ij}-n_{i}n_{j},
\end{align}
$\mathcal{Q}^{ij}$ is the mass quadrupole moment
\begin{align}
\nonumber
\mathcal{Q}^{ij}&=I^{ij}-\frac{1}{3}\delta^{ij}I^{kk},\\
I^{ij}&=\frac{1}{c^2}\int T^{00}(\vec{x},t) x^{i}x^{j}\mathrm{d}x^3,
\end{align}
$\mathcal{O}^{ijk}$ is the mass octupole moment
\begin{align}
\nonumber
\mathcal{O}^{ijk}&=I^{ijk}-\frac{1}{5}\left[\delta^{ij}I^{llk}+\delta^{ik}I^{ljl}+\delta^{jk}I^{ill}\right],\\
I^{ijk}&=\frac{1}{c^2}\int T^{00}(\vec{x},t)x^{i}x^{j}x^{k}\mathrm{d}x^3,
\end{align}
and $\mathcal{Z}^{ijk}$ is the current quadrupole moment
\begin{align}
\nonumber
\mathcal{Z}^{ijk}&=I^{i,jk}+I^{j,ik}-2I^{k,ij},\\
I^{i,jk}&=\frac{1}{c}\int T^{0i}(\vec{x},t)x^{j}x^{k}\mathrm{d}x^3.
\end{align}
Given the energy-momentum tensor $T_{\mu\nu}$ during simulations, we give two GW polarizations explicitly for GWs traveling in the z-direction ($n^{i}=(0,0,1)$)
\begin{align}
\nonumber
h_{+}&=h_{11}^{TT}(t-r_0/c),\\
h_{\times}&=h_{12}^{TT}(t-r_0/c),
\end{align}
where the distance from source $r_0$ is set as $1$Mpc in Fig.~\ref{fig:gw1}, Fig.~\ref{fig:gw2} and Fig.~\ref{fig:gw3}. 
In these figures, we demonstrate that the contributions from $\mathcal{Q}^{ij}$ dominate over those from $\mathcal{O}^{ijk}$. We also confirm that the contributions from $\mathcal{Z}^{ijk}$ are even much smaller than those from $\mathcal{O}^{ijk}$. Therefore, the following discussion only takes the contributions from $\mathcal{Q}^{ij}$ into consideration.
Since our simulations are completely independent of the FDM mass $m$, we can rescale the amplitude and frequency of GWs by fine-tuning $m$ according to Eq.~(\ref{eq:unit}) and Eq.~(\ref{eq:dless2}). For example, the GW frequency is typically (few ten-years)$^{-1}$ for $m=10^{-18}\rm{eV}/c^2$ or (few years)$^{-1}$ for $m=10^{-17}\rm{eV}/c^2$.
Also, from these three figures, we find that the $h_{\times}$ polarizations are zero, which naturally results from a head-on collision in the $x$ direction, and the $h_{+}$ polarizations have a frequency higher than the frequency of energy transfer between $E_{\rm{p}}$ and $E_{\rm{k}}$, as shown in the left subplots of Fig.~\ref{fig:C1}, Fig.~\ref{fig:C2} and Fig.~\ref{fig:C3}. 

To validate the neglect of the GW back reaction on the evolution of FDM solitons during our simulations, we integrate the energy carried away by GWs during the whole simulation and compare this integration $E_{\rm{\rm{GW}}}$ with the energy transfer between $E_{\rm{p}}$ and $E_{\rm{k}}$ in a period. We find that the former one is
\begin{equation}
E_{\rm{\rm{GW}}}=\frac{c^3r^2}{32\pi G}\int d\Omega\int dt\frac{\partial h^{TT}_{ij}}{\partial t}\frac{\partial h_{TT}^{ij}}{\partial t},
\end{equation}
which is much smaller than the latter one, such as $E_{\rm{\rm{GW}}}/\Delta E_{\rm{p}}\approx 3\times10^{-12}$ for C1, \Ke{$E_{\rm{\rm{GW}}}/\Delta E_{\rm{p}}\approx 4\times10^{-13}$} for C2 and \Ke{$E_{\rm{\rm{GW}}}/\Delta E_{\rm{p}}\approx 9\times10^{-13}$} for C3. Therefore, our approximation to the GW back reaction is reasonable.

Rewriting the total energy $E_{\rm{\rm{GW}}}$ in the frequency domain, the energy spectrum can be calculated for every collision, as shown by these blue curves in Fig.~\ref{fig:dedf},
\begin{eqnarray} 
\label{eq:dedf1}
\frac{dE_{\rm{GW}}}{df}=\frac{\pi c^{3}r^{2}f^{2}}{2G}\int d\Omega  \left ( \left |\tilde{h}_{+}(f)\right |^{2}+\left |\tilde{h}_{\times}(f)\right |^{2}\right ).
\end{eqnarray}
For collisions between two FDM solitons both with the density profile as the ground state, we construct a phenomenological model of $\frac{dE_{\rm{GW}}}{df}$, which is calibrated to $\frac{dE_{\rm{GW}}}{df}$ of C1 and C2, as shown by these red curves in Fig. 8,
\begin{align}
\label{eq:dedf2}
\nonumber
\frac{G}{c^5}\frac{dE_{\rm{GW}}}{df}=&s^2\times\left(\frac{\lambda_1}{1.66\times 10^{-6}}\right)^3\times
10^{-2\left(\frac{1+q}{q}\right)^{\Ke{\frac{3}{25}}}(1-q)}\times\\
&\sum_{i=1}^5a_{i}e^{-\left(\frac{\frac{f}{[\rm{Hz}]}\Ke{\frac{(1+2q)s}{3q}}-b_i}{c_i}\right)^2}{\rm Hz^{-2}},
\end{align}
where the values of $15$ phenomenological parameters $\{a_i,b_i,c_i\}$ are listed in Tab.~\ref{tb:coefficients} and $\{s,q,\lambda_1\}$ are $3$ free parameters: $s=\frac{10^{-18}\rm{eV/c^2}}{m}$, $q\le1$ is the mass-ratio between two FDM solitons and $\lambda_1$ is the scaling factor of the more massive FDM soliton.
If above model of $\frac{dE_{\rm{GW}}}{df}$ is applicable over a large enough region of parameter space $\{s,q,\lambda_1\}$, we can obtain the corresponding stochastic GW background (SGWB) energy-density spectrum~\cite{Phinney:2001di}
\begin{align}
\nonumber
\Omega_{\rm{GW}}(m)=\frac{8\pi G f}{3H_{0}^2c^2}\int dzd\lambda_1dq\frac{dM_{\rm{halo}}}{d\lambda_1}\frac{dq_{\rm{halo}}}{dq}\\\frac{d^3n_{\rm{halo}}}{dzdM_{\rm{halo}}dq_{\rm{halo}}}\frac{dE_{GW}}{df_{s}},
\end{align}
where $f_s=f(1+z)$ is the frequency in the source frame, $H_0=67.36{\rm km/s/Mpc}$ is the Hubble constant~\cite{Planck:2018vyg}, $q_{\rm{halo}}$ is related to $q$ according to Eq.~(\ref{eq:halo0}), $M_{\rm{halo}}$ is related to $\lambda_1$ as Eq.~(\ref{eq:halo1}) and $\frac{d^3n_{\rm{halo}}}{dzdM_{\rm{halo}}dq_{\rm{halo}}}$ is the halo differential merger rate per unit redshift, halo mass and mass ratio~\cite{Bi:2023tib}.
Finally, we compare the GW signals from FDM soliton collisions with the sensitivity curves of some PTA experiments (adapted from~\cite{Sesana:2008mz,Khmelnitsky:2013lxt,DeMartino:2017qsa}) in Fig.~\ref{fig:pta}, where $\Omega_{\rm{GW}}$ and $|\tilde{h}(f)|$ are related to the characteristic strain $h_c$ as $H_0^2\Omega_{\rm{GW}}=\frac{2\pi^2}{3} f^2h_c^2$ and $2f|\tilde{h}(f)|=h_c(f)$ respectively~\cite{Moore:2014lga}. Due to the lower merger rate for massive halos $M_{\rm{halo}}\gtrsim10^{13}\rm{M}_{\odot}$, the characteristic strain predicted from the SGWB by all FDM soliton collisions is much lower than the sensitivity curves.

\begin{table}[htbp]
\centering
\caption{Values of $18$ phenomenological parameters.}
\label{tb:coefficients}
\Ke{
\begin{tabular}{|c|ccc|}
\hline 
 $i$ & $a_i$ & $b_i$  & $c_i$  \\
\hline 
1 & $7.105 \times 10^{45}$ & $1.298 \times 10^{-9}$ & $4.222 \times 10^{-10}$ \\
2 & $1.287 \times 10^{45}$ & $1.949 \times 10^{-9}$ & $4.816 \times 10^{-10}$ \\
3 & $3.498 \times 10^{43}$ & $9.410 \times 10^{-10}$ & $6.093 \times 10^{-10}$ \\
4 & $3.292 \times 10^{44}$ & $2.827 \times 10^{-9}$ & $6.339 \times 10^{-10}$ \\
5 & $5.601 \times 10^{43}$ & $3.541 \times 10^{-9}$ & $7.519 \times 10^{-10}$ \\
6 & $7.258 \times 10^{43}$ & $4.810 \times 10^{-9}$ & $8.682 \times 10^{-10}$ \\
\hline 
\end{tabular}
}
\end{table}

\begin{figure*}[]
\begin{center}
\subfloat{\includegraphics[width=1.0\textwidth]{gwc3.png}}
\end{center}
\captionsetup{justification=raggedright}
\caption{GWs with $h_{+}$ polarization (blue curves) and $h_{\times}$ polarization (red lines) from C3, where the distance from source is $1$Mpc and the FDM mass is $m=10^{-17}\rm{eV}/c^2$ (left) or $m=10^{-18}\rm{eV}/c^2$ (right).}
\label{fig:gw3}
\end{figure*}

\begin{figure*}[]
\begin{center}
\subfloat{\includegraphics[width=1.0\textwidth]{dedf.png}}
\end{center}
\captionsetup{justification=raggedright}
\caption{Energy spectra in the source frame for C1 (upper two subplots) and C2 (lower two subplots), which can be calculated directly from simulation data with Eq.~(\ref{eq:dedf1}) (blue curves) or fitted with a parameterized formula Eq.~(\ref{eq:dedf2}) (red curves).}
\label{fig:dedf}
\end{figure*}

\begin{figure*}[]
\begin{center}
\subfloat{\includegraphics[width=1.0\textwidth]{pta.png}}
\end{center}
\captionsetup{justification=raggedright}
\caption{Comparison between the sensitivity curves of some PTA experiments (adapted from~\cite{Sesana:2008mz,Khmelnitsky:2013lxt,DeMartino:2017qsa}) with characteristic strains predicted from only C1 with $m=10^{-17}\rm{eV}/c^2$ (red solid) or $m=10^{-18}\rm{eV}/c^2$ (red dashed) or the SGWB by all FDM soliton collisions with $m=10^{-17}\rm{eV}/c^2$ (blue solid) or $m=10^{-18}\rm{eV}/c^2$ (blue dashed) respectively.}
\label{fig:pta}
\end{figure*}

\section{Summary and Discussion}
\label{sec:sd}
In this paper, we propose the generation of GWs from collision of FDM solitons. Firstly, we turn to the shooting method to solve the SP system with spherical symmetry and build up the density profiles of the ground state and the first excited state of an isolated FDM soliton in Section~\ref{sec:soliton1}. In this section, we also propose a new adimensional unit of length, mass and time. 
In Section~\ref{sec:soliton2}, according to these new adimensional units, we simulate three head-on collisions: two solitons with the mass ratio equal to $1$ or $0.5$ and with the density profile as the ground state; two solitons with the mass ratio equal to $0.5$ but with the density profile as the ground state and the first excited state respectively. We find that the gravitational potential is strong enough to merge the initial two FDM solitons with each other and, during collisions, the energy transfer between the kinetic energy and the gravitational potential energy causes the final FDM solitons to oscillate irregularly with an adimensional frequency of $\sim10^{-8}$. As shown in Section~\ref{sec:gw}, due to this spherically asymmetric oscillation, GWs are emitted and can be easily calculated when the linearized theory is valid and the GW back reaction on the evolution of FDM solitons is ignored. By fine-tuning FDM mass $m=10^{-18}\rm{eV}/c^2$ or $m=10^{-17}\rm{eV}/c^2$, the GWs from collisions have a frequency of (few ten-years)$^{-1}$ or (few years)$^{-1}$ respectively.
However, their own characteristic strains or the corresponding SGWB's characteristic strains are out of the current and near future detection.

In contrast to the adimensional unit of length, mass and time applied in~\cite{Paredes:2015wga,Edwards:2018ccc,Schive:2014hza}, namely $\mathcal{L}\equiv\left(\frac{8\pi\hbar^2}{3 m^2H_0^2\Omega_{m}}\right)^{1/4}\approx121\left(\frac{10^{-23}\rm{eV}/c^2}{m}\right)^{1/2}\rm{kpc}$, $\mathcal{M}\equiv\frac{1}{\rm{G}}\left(\frac{8\pi}{3 H_0^2\Omega_{m}}\right)^{-1/4}\left(\frac{\hbar}{m}\right)^{3/2}\approx 7\times 10^7\left(\frac{10^{-23}\rm{eV}/c^2}{m}\right)^{3/2}\rm{M}_{\odot}$ and $\mathcal{T}\equiv\left(\frac{8\pi}{3 H_0^2\Omega_{m}}\right)^{1/2}\approx75.5 \rm{Gyr}$, our scales (Eq.~(\ref{eq:unit})) are more suitable to study the evolution of FDM solitons in years. For example, the physical velocity $\vec{v}=\tilde{\vec{v}}\mathcal{L}\mathcal{T}^{-1}$ in this paper is independent of FDM mass $m$ and we can easily fix it as $\left(10^{-4}c,0,0\right)$. As a result, the simulations are completely independent of $m$ and the dynamics of FDM solitons can be simply rescaled by adjusting $m$. However, the dimensionless velocity $\tilde{\vec{v}}$ \Mov{does} depend on $m$ in~\cite{Paredes:2015wga,Edwards:2018ccc,Schive:2014hza} if one fixes the physical velocity $\vec{v}=\tilde{\vec{v}}\mathcal{L}\mathcal{T}^{-1}$ as $\left(10^{-4}c,0,0\right)$ also. On the other hand, the dimensionless velocity $\tilde{\vec{v}}$ is an essential parameter which must be provided as one of the initial conditions. Therefore, any adjustment about $m$ will affect $\tilde{\vec{v}}$, which means that the corresponding simulation must be run again and it is not efficient enough to scan the whole parameter space and pick out the more physical systems.

The head-on collision may be not realistic enough for the evolution of FDM solitons. Usually, there should be some angular momentum for the system of FDM solitons even though it is difficult for FDM solitons to form a regular binary system because of their large size and weak self-gravitational potential sourced by their FDM density.
Consequently, the presence of some angular momentum would almost balance the energy carried away by different GW polarization. 
\Mov{For the late evolution of a bound system coming from a two-soliton system with some angular momentum, we can approximate it with a post-collision as well as a rotation.}
Therefore, we suppose that our results on the frequency and amplitude of GWs from collision of FDM solitons are also reasonable for other more general evolution of FDM solitons.

\begin{acknowledgments}
Ke Wang is supported by grants from the National Key Research and Development Program of China (Grant No. 2021YFC2203003). Chen Tan and Jing-Kang Bin is supported by grants from NSFC (Grant No.12247101).
\end{acknowledgments}

\end{document}